\documentclass[epj]{svjour}
\usepackage{graphicx}
\begin{document}
\title{Community analysis in social networks}
\author{Alex Arenas\inst{1} \and Leon Danon\inst{2} \and Albert D\'{\i}az-Guilera\inst{2}
\and Pablo M. Gleiser \inst{2} \and Roger Guimer{\`a} \inst{3}
%
}                     
%
%
\institute{Departament d'Enginyeria Inform{\`a}tica i
Matem{\`a}tiques, Universitat Rovira i Virgili, Avda. dels Paisos
Catalans, 43006 Tarragona, Spain \and Department de F\'{\i}sica
Fonamental, Universitat de Barcelona, Mart\'{\i} i Franqu\`{e}s 1,
08028 Barcelona, Spain \and Department of Chemical Engineering,
Northwestern University, Evanston, IL, USA}
\date{Received: date / Revised version: date}

\abstract{We present an empirical study of different social networks
obtained from digital repositories. Our analysis reveals the community
structure and provides a useful visualising technique.  We investigate
the scaling properties of the community size distribution, and that
find all the networks exhibit power law scaling in the community size
distributions with exponent either $-0.5$ or $-1$. Finally we find
that the networks' community structure is topologically self-similar
using the Horton-Strahler index.  
} 
\maketitle
\section{Introduction}
\label{sec:intro}

The topology of complex networks have been the subject of intensive
study over the past few years. It has been recognised that such
topologies play an extremely important role in many systems and
processes, for example, flow of data in computer networks
\cite{Menezer03}, energy flow in food webs \cite{Garlasch03},
diffusion of information in social networks \cite{newman_rev},
etc. This has led to advances in fields as diverse as computer
science, biology and social science to name but a few.

It has recently been found that social networks exhibit a very clear
community structure. For example, in an organisation, such community
structure corresponds, to some extent, to the formal chart, and to
some extend to ties between individuals arising due to personal,
political and cultural reasons, giving rise to informal communities
and to an {\it informal community structure}. The understanding of
informal networks underlying the formal chart and of how they operate
are key elements for successful management. In other scenarios, this
community structure reflects in general the self-organisation of
individuals to optimise some task performance, for example, optimal
communication pathways or even maximisation of productivity in
collaborations. Characterising and understanding this structure may be
fundamental to the study of dynamical processes that occur on these
nets. In this paper we present the empirical study of several social
networks at the level of community structure. We show that all exhibit
self-similar properties, with the community size distributions
following power laws. The exponents of these power laws seem to fall
into two distinct classes, one with exponent $\sim -0.5$ and the
other with exponent $\sim -1$. The source of these two different
scaling laws is still being investigated.

In the next section we describe the methodology used to characterise
the social structure of the networks we study. In Section
\ref{sec:applications}, we apply this methodology to various networks,
and in Section \ref{sec:char} we characterise the community structure.
Finally we present an interpretation of the results and propose some
future work.

\section{The method}
\label{sec:method}
\subsection{Identification of real communities}

The traditional method for identifying communities in networks is
hierarchical clustering \cite{jain88}.  Given a set of N nodes to be
clustered, and an N$\times$N distance (or similarity) matrix, the basic
process of hierarchical clustering is this: Start by assigning each
node its own cluster, so that if you have N nodes, you now have N
clusters, each containing just one node. Let the distances between the
clusters equal the distances between the nodes they contain. Find the
closest (or most similar) pair of clusters and merge them into a
single cluster, so that now you have one less cluster. Compute
distances between the new cluster and each of the old clusters. Repeat
until all nodes are clustered into a single cluster of size N. 

\begin{figure}
\centerline{\includegraphics*[width=\columnwidth]{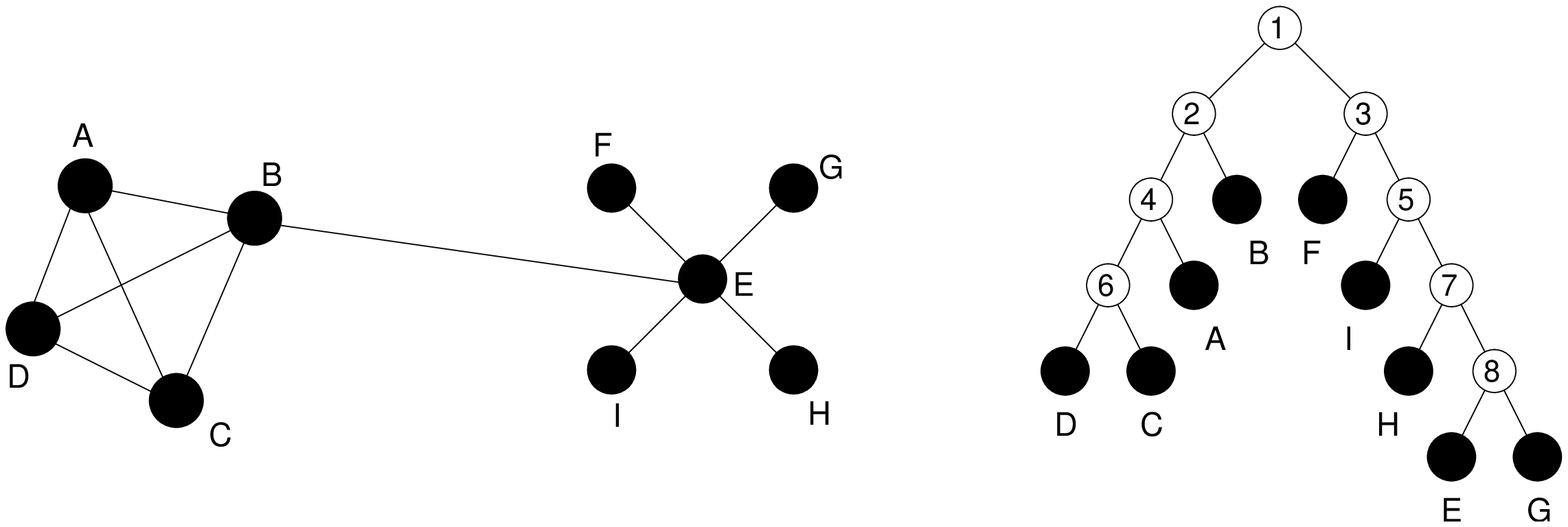}}
\centerline{(a)\hspace{0.5\columnwidth}(b)}
\caption{Community identification according to the GN algorithm. (a) A
network containing two clearly defined communities connected by the
link $BE$. This link will have the highest betweenness, since to get
from any node in one community, to any node in the other, this link
needs to be used. Therefore it will be the first link to be cut,
splitting the network in two. The process of cutting this link
corresponds to the bifurcation at the highest level of the binary tree
in (b). Since there is no further community structure in the offspring
networks, the rest of the nodes will be separated one by one,
generating a binary tree with two branches corresponding to the two
communities. For the community on the right, the most central node
will be separated last. In general, branches of the binary tree
correspond to communities of the original network and the tips of
these branches correspond to the central nodes of the communities.}
\label{algorithm}
\end{figure}

In this work we use a different community identification algorithm,
proposed recently by Girvan and Newman (GN) \cite{girvan02}. This new
algorithm gives successful results even for networks in which
hierarchical clustering methods fail. The algorithm works as
follows. The betweenness of an edge is defined as the number of
minimum paths connecting pairs of nodes that go through that edge
\cite{wasserman94,newman01}.  The GN algorithm is based on the idea
that the edges which connect highly clustered communities have a
higher edge betweenness---for example edge $BE$ in Figure
\ref{algorithm}a---and therefore cutting these edges should separate
communities. Thus, the algorithm proceeds by identifying and removing
the link with the highest betweenness in the network. This process is
repeated (should it be necessary) until the `parent network splits,
producing two separate `offspring' networks. The offspring can be
split further in the same way until they contain only one
node. In order to describe the entire splitting process, we
generate a binary tree, in which bifurcations (white nodes in Figure
\ref{algorithm}b) depict communities and leaves (black nodes)
represent individuals. All the information about the community
structure of the original network can be deduced from the topology of
the binary tree constructed in this fashion.

\subsection{Graphical representation of the hierarchical community structure}\label{graph}

Consider again the network in Figure \ref{algorithm}a. At the
beginning of the process, no links have been removed and the whole
network is represented by node 1 in the binary tree of Figure
\ref{algorithm}b. When edge $BE$ is removed, the network splits in two
groups: group 2, containing nodes $A$ to $D$, and group 3, containing
nodes $E$ to $I$. After this first splitting, two completely separate
communities are left, a very homogeneous one and a very centralised
one. One can check that in both cases the algorithm will separate
nodes one by one giving rise to two different branches in the binary
tree. Actually, when communities with no further internal structure
are found, they are disassembled in a very uneven way giving rise to
branches. In other words, the almost impossible task of identifying
communities from the original network is replaced by the easy task of
identifying branches in the binary tree. When centralised network
structures are treated, the central node(s) will appear at the end of
the branch, thus also providing a method of identifying the
``leaders'' of each community.

\section{Applications}
\label{sec:applications}

In this section we apply the method described in the previous section
to various networks. In Table 1 we present the characterising
statistics of each of the networks.
\begin{table}[!ht]
\centering
{\begin{tabular}{|c|c|c|c|}
\hline
Network & $N$ & $\langle d\rangle$ & $\langle C\rangle$\\ 
\hline
mail & 1134 & 2.42 & 0.31\\ 
jazz & 1265 & 2.79 & 0.89\\
fises & 784 & 5.71 & 0.78\\
gr-qc & 2546 & 6.11 & 0.54\\
hep-lat & 1411 & 4.71 & 0.66\\
quant-ph & 1460 & 5.97 & 0.71\\
math-ph & 2117 & 10.13 & 0.58\\
\hline
\end{tabular}}
\label{table:standard}
\caption{Statistics for the networks we study. $N$ is the number of
nodes in the network,$\langle d\rangle$ is the average distance
between nodes and $\langle C\rangle$ is the clustering
coefficient. Note that the clustering coefficient all the networks
apart from the mail network are extremely high. This is due to the
networks' construction as bipartite graphs. For example, in the ArXiv
network, if four authors coauthor only one paper, the clustering
coefficient of those four nodes in the network will be 1.}
\end{table}
\subsection{E-mail network}

We extract and build a network of interactions via e-mail using logs
from mail servers over a period of 3 months. In order to be able to
concentrate on the real social structure, we remove 'spam' mails with
more than 50 recipients, and only create links between people that
have {\it exchanged} e-mails, that is, an e-mail that was sent from A
to B was responded to within the 3 month period. More information can
be found in \cite{guimera??b}.

\begin{figure}[h!]
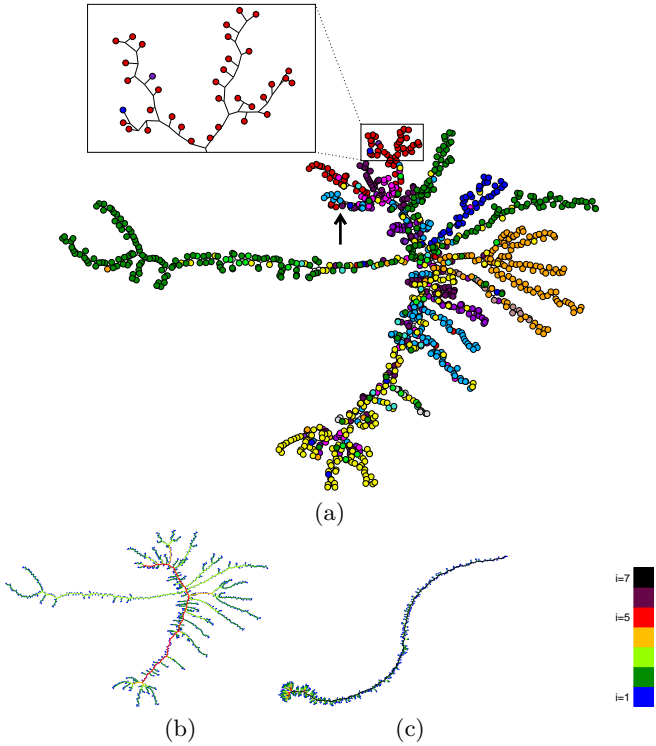

\begin{minipage}[]{\columnwidth}
\centerline{\includegraphics*[width=0.82\columnwidth]{fig3a}}
\centerline{(a)}
\centerline{\includegraphics*[width=0.4\columnwidth]{hot-cold}
\includegraphics*[width=0.35\columnwidth]{randhot-cold}
\quad\quad\quad\quad\includegraphics*[width=0.06\columnwidth]{legend}}
\hspace{0.25\columnwidth}(b)\hspace{0.3\columnwidth}(c)
\caption{(a) Binary tree showing the result of applying the GN
algorithm and our visualisation technique to the e-mail network of
URV. Each branch corresponds to a real community and the tips of the
branches correspond to their leaders. The splitting procedure starts
in the position indicated by an arrow at the top of the drawing and
proceeds downward. The colour of the nodes represents different centres
within the university (five small centres containing less than 10
individuals are assigned the same colour). Nodes of the same colour
(from the same centre) tend to stick together meaning that individuals
within the same centre tend to communicate more, and that the
algorithm is capable of resolving separate centres to a good degree of
accuracy. (b) Same as before but without showing the nodes, so that
the structure of the tree is clearly shown. Branches are coloured
according to their Horton-Strahler index (see Section \ref{sec:HS})
(c) Binary tree showing the result of applying the GN algorithm to a
random graph with the same size and degree distribution than the
e-mail network. Again, colours correspond to Horton-Strahler indices.}
\label{arbre}
\end{minipage}
\end{figure}

Figure \ref{arbre}a shows the binary tree that results from the
application of our method to the e-mail network of URV.  Each colour
corresponds to an individual's affiliation to a specific centre within
the university. Centres are in most of the cases faculties or
colleges---for example the School of Engineering---and are usually
comprised of departments---for example, the Department of Computer
Sciences and Mathematics or the Electrical Engineering Department. In
turn, departments are divided into research teams---for instance, the
group of Complex Systems or the group of Dynamical Systems in the
Department of Computer sciences and Mathematics.

Instead of plotting the binary tree with the root at the top as in
Figure \ref{algorithm}b, it is plotted optimising the layout so that
branches, that represent the real communities, are as clear as
possible. Actually, the root is located at the position indicated with
the arrow in the upper left region of the tree.  The branches obtained
by the GN procedure (Figure \ref{arbre}) are essentially of one colour,
indicating that we have correctly identified the centres of the
university. This is especially true if one focuses on the ends of the
branches since, as discussed above, these ends correspond to the most
central nodes in the community. In regions close to the origin of the
branches, the coexistence of colours corresponds to the boundary of a
community. It is important to note that the GN algorithm is able to
resolve not only at the level of centres, but is also able to
differentiate groups (sub-branches) inside the centres, i.e.,
departments and even research teams.

For comparison, we also show the tree generated by the GN algorithm
from a random graph of the same size and degree distribution as the
e-mail network (Figure \ref{arbre}c). The absence of community
structure is apparent from the plot.

\subsection{The Jazz network}

In this section we construct and study the network of jazz musicians
 obtained from the Red Hot Jazz Archive of recordings between 1912 and
 1940 (www.redhotjazz.com), at two different levels.  First we build
 the network from a 'microscopic' point of view.  In this case each
 vertex corresponds to a musician, and two musicians are connected if
 the have recorded in the same band. Then we build the network from a
 'coarse-grained' point of view. In this case each vertex corresponds
 to a band, and a link between two bands is established if they have
 at least one musician in common. This is the simplest way in which one
 can establish a connection between bands, and the definition can be
 extended to incorporate directed and/or weighted links. However, we
 show that even by using this simple definition we are able to recover
 essential elements of the community structure. More information can
 be found in \cite{gleiser03}.

\begin{figure}
   \centerline{\includegraphics[width=0.7\columnwidth]{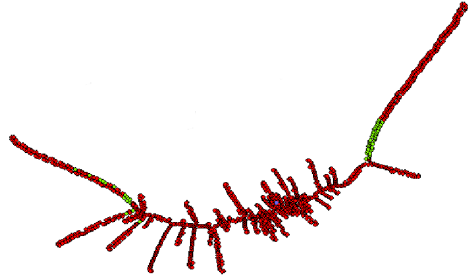}}
   \vspace*{8pt}
   \caption{\label{jazz_1} Community structure of the jazz musicians
     network. The root of the tree, in the middle of the figure, is
     indicated with the colour blue.  The musicians with $k>170$ are
     indicated with green.}
 \end{figure}

In Figure \ref{jazz_1} we show the binary tree corresponding to the
 musicians network. The root of the tree is indicated with a blue
 circle. A clear separation into two distinct communities can be can be
 seen and can be interpreted as the manifestation of racial
 segregation present at that time. Although a small number of
 collaborations existed between races, most bands were exclusively
 comprised of one race or the other.  As a consequence a division in
 two large communities separating black and white musicians should be
 present. In fact, an analysis of the names of the musicians shows
 that the musicians on the left community are black while the
 musicians on the right are white.
 As in the e-mail network, the most central musicians are expected to
 appear at the end of the branches. However in Figure \ref{jazz_1} we see
 that those musician with $k>170$ appear at the beginning of the
 branches. This appears to be an artefact of the manner in which the
 network is created, as these musicians {\it must} have played in more
 than one band. Therefore, their affiliation with the rest of the
 musicians in the branch they appear in is relatively lower. Also,
 since everyone plays with everyone in a band, there is {\it no} well
 defined central Figure.

\begin{figure}
  \includegraphics[width=0.7\columnwidth]{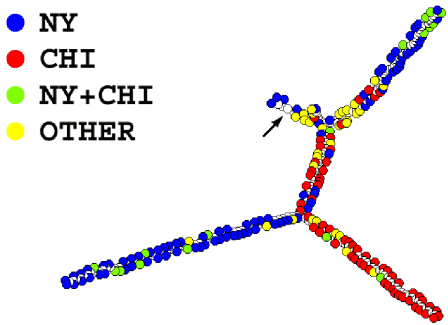}
  \caption{\label{jazz_2} Communities in the jazz bands network. The arrow
    indicates the root of the tree.  The different colours correspond to
    cities where a band has recorded: New York (blue), Chicago (red), both
    in New York and Chicago (green) and other cities (yellow).}
\end{figure}

A similar effect can be seen when analysing the bands network. The
binary tree shown in Figure \ref{jazz_2} reveals a very simple community
structure. The tree is roughly divided into two large communities as
expected. However, the largest branch also splits into two. To
understand the origin of this division we have analysed the cities
where the bands recorded. We indicate with colour red the bands that
have recorded in New York. The bands that recorded in Chicago are
indicated with blue.

In this case, central bands do play an important role. The analysis of
names \cite{grove} shows that the bands at the tip of the branches
were the some of the most influential in the epoch. In general they
also contained the most connected musicians.

These results show that both the musicians and bands network capture
 essential ingredients of the collaboration network of jazz
 musicians. 

\subsection{FisEs}

We construct a network of scientists that contributed to the
Statistical Physics (F\'{\i}sica Estad\'{\i}stica) conferences in
Spain over the last 16 years. In a similar approach to the one
described below, we consider two scientists linked if they have
co-authored a panel contribution to any of the conference. To be able
to consider the historical structure of this network we ``accumulate''
the network over all the conferences, that is, once a link is created,
it remains, even if the authors never collaborated again. The final
network (accumulated over all the years) is comprised of 784 nodes
with 655 (84\%) of those belonging to the giant component.

In the figure below we show the binary tree as generated by our
formalism. The colours in this case represent the universities or
centres of investigation of the participants. Those nodes whose
affiliation has not been identified and those that belong to
institutions outside of Spain are not shown, since they are few, and
do not play an important role in the structure of this network. The
colours in the figure represent the centres of origin of the
contributors have been identified, and the grey
nodes represent all universities with just a few contributions.

\begin{figure}[h!]
\centerline{\includegraphics*[width=0.82\columnwidth]{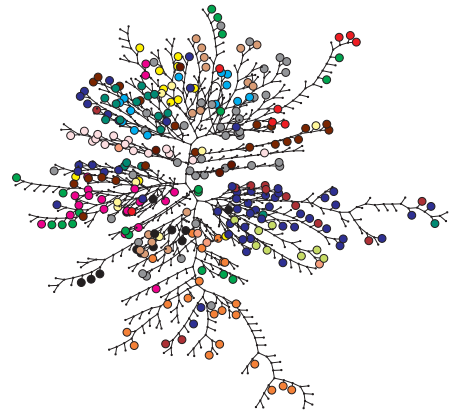}}
 \caption{Binary tree showing the result of applying the GN algorithm
 and our visualisation technique to the network of coauthors in
 FisEs. Each branch corresponds to a real community and the tips of
 the branches correspond to the people that have played a major role
 in the different research groups. Nodes of the same colour (from the
 same centre) show up mainly in the same branches, showing that
 collaborations are more common within centres than between them.}
\label{fises_tree}
\end{figure}

\subsection{arXiv}

Finally, we study the community structure of the network of scientific
collaborations as extracted from xxx.arxiv.org preprint repository
\cite{newman01a}. Scientists are considered linked if they have
coauthored a paper in the repository. The articles defining the links
are classified into different fields. Due to the size of the entire
network (52909 nodes, 44337 of which are connected in a giant cluster)
we create $4$ separate networks, each corresponding to one of the
following fields: Mathematical Physics(math-ph), High Energy Physics -
Lattice (hep-lat), General Relativity and Quantum Cosmology (gr-qc),
Quantum Physics (quant-ph).  An extensive study of the geographic
location and thematic affiliations of the authors has not yet been
performed. 

\begin{figure}
\centerline{\includegraphics*[width=0.43\columnwidth]{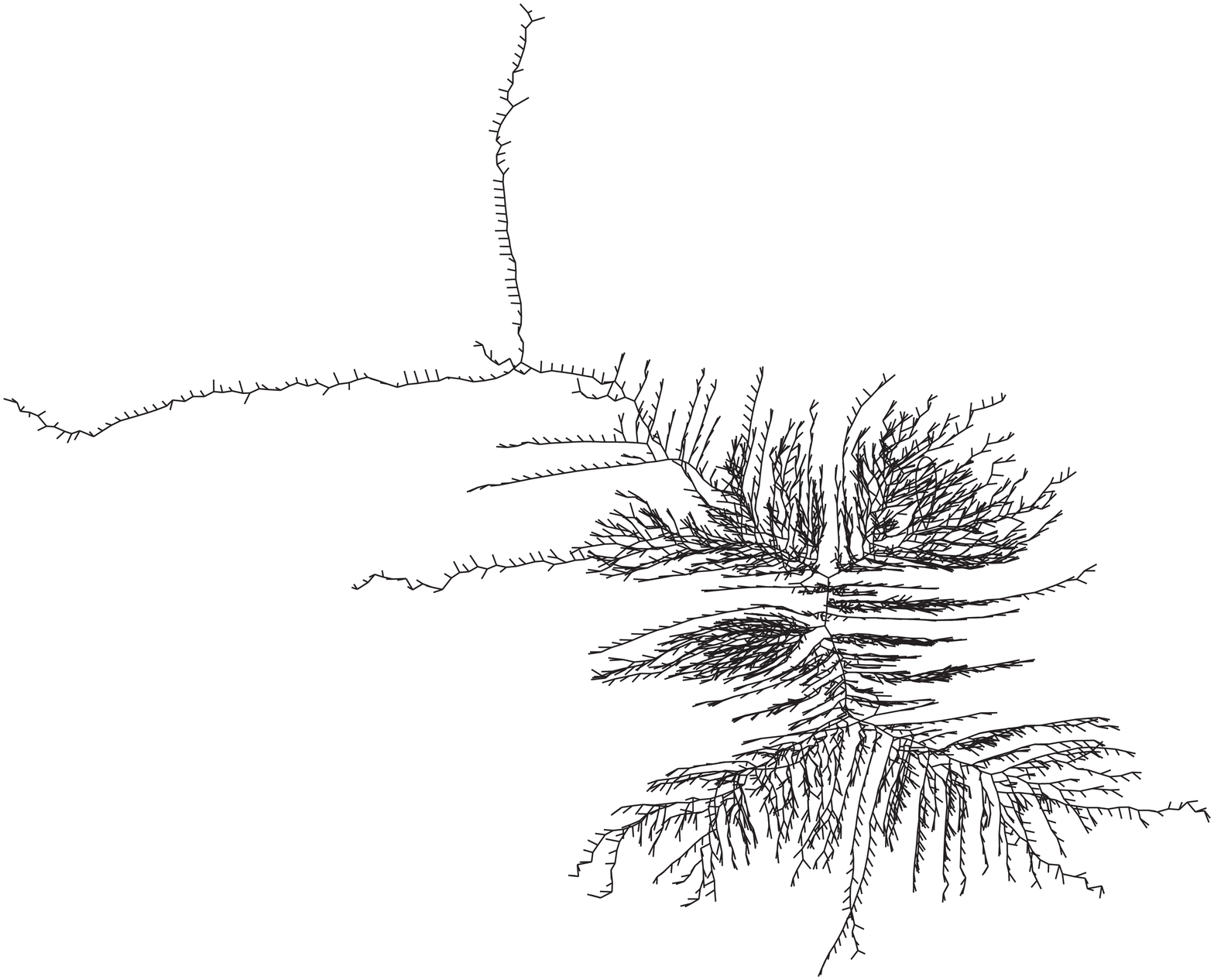}
            \includegraphics*[width=0.43\columnwidth]{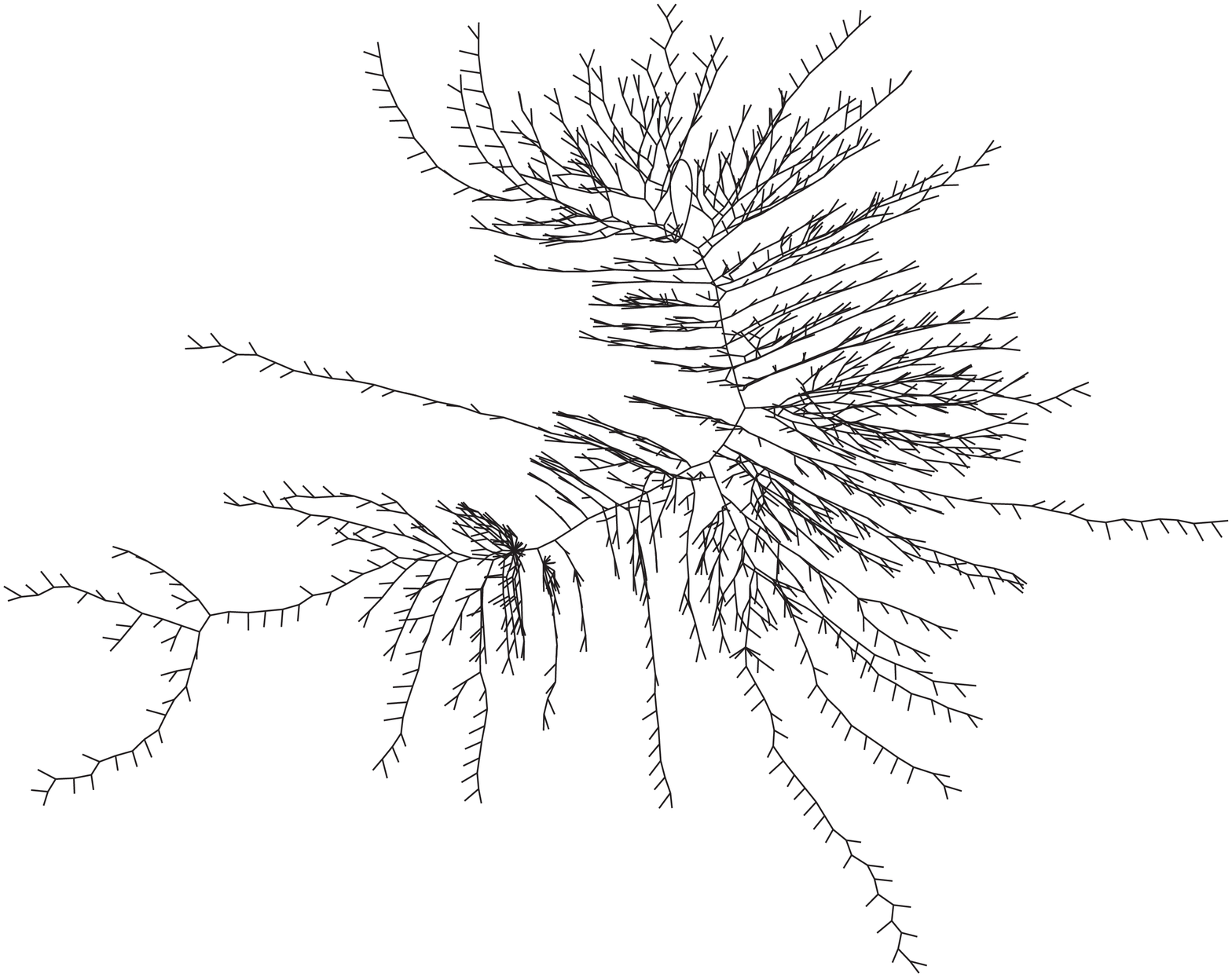}}
\centerline{gr-qc\hspace{0.43\columnwidth}quant-ph}
\centerline{\includegraphics*[width=0.43\columnwidth]{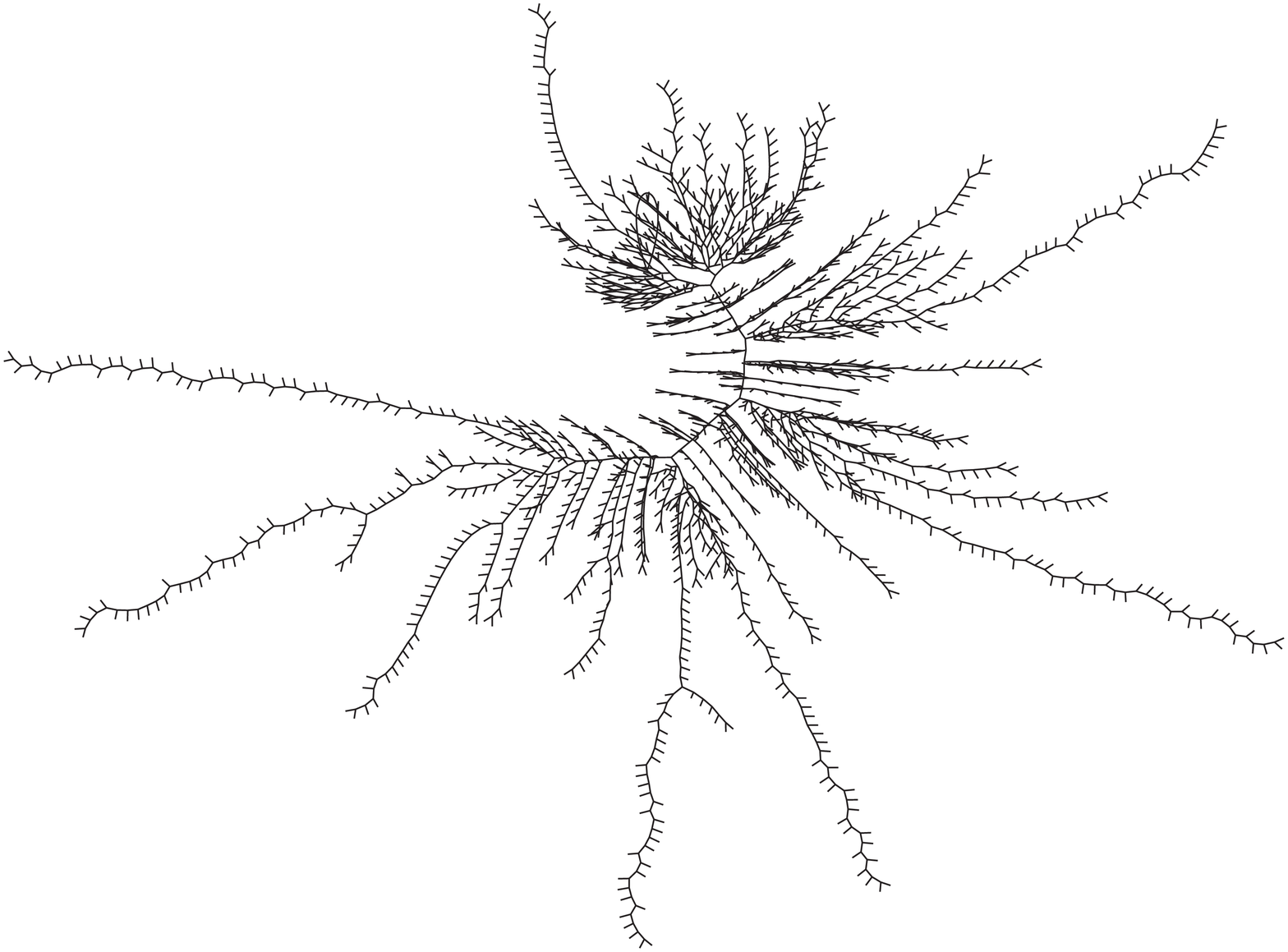}
            \includegraphics*[width=0.43\columnwidth]{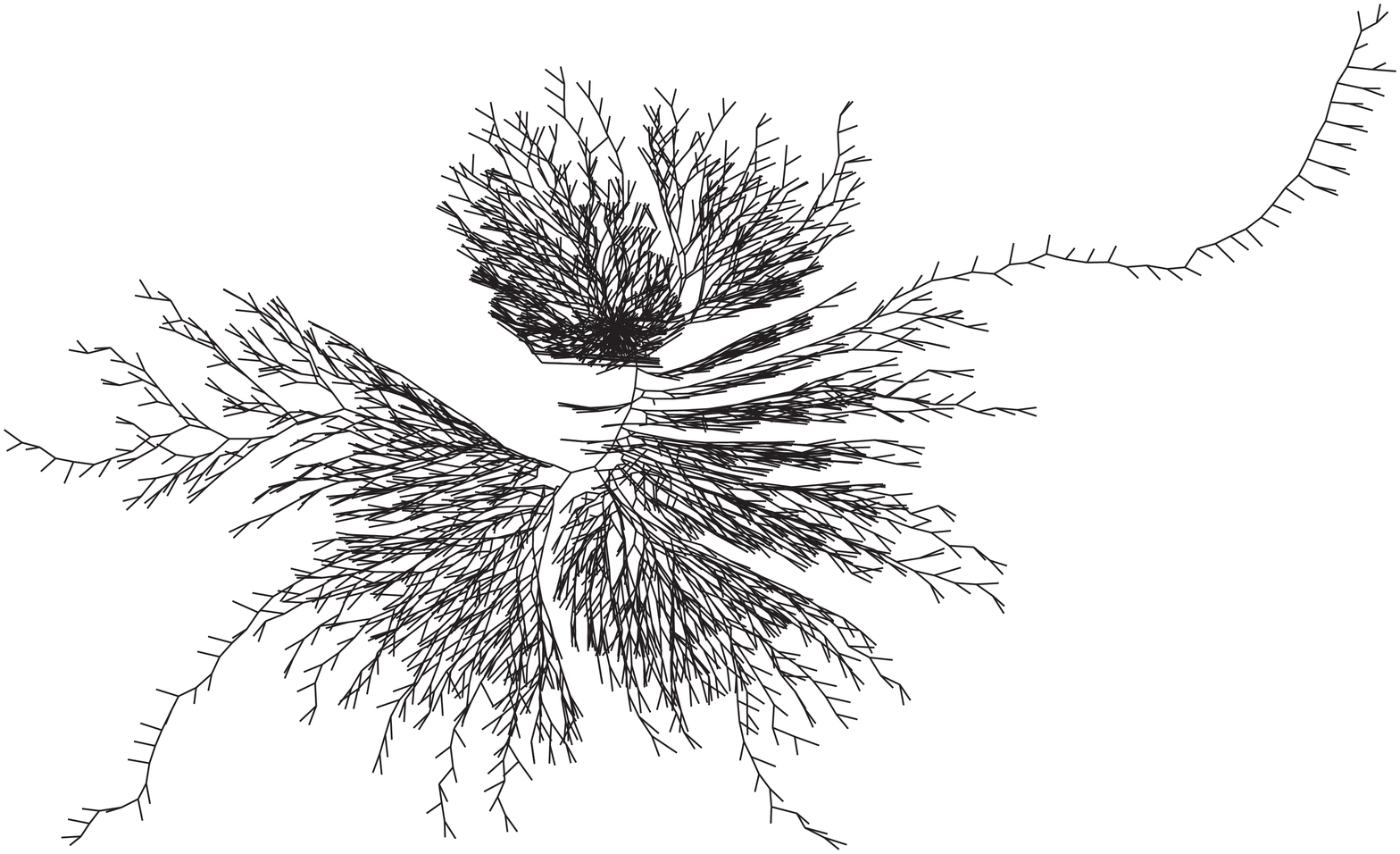}}
\centerline{hep-lat\hspace{0.43\columnwidth}math-ph}
\caption{Binary trees showing the results of the community separation
applied to four different parts of the arXiv network. Clear community
structure is once again seen, probably corresponding to different
paper themes and interests of authors, as well as geographic location.}
\label{archiv_tree}
\end{figure}

\section{Emergent properties of the community structure}
\label{sec:char}
In this section we characterise the statistical properties of the
community structure of the networks analysed in the previous section.
We will show that there are self similar properties that emerge in the
network community structure.

\subsection{Community size distribution}

\begin{figure}
\begin{minipage}[]{\columnwidth}
\centerline{\includegraphics*[width=\columnwidth]{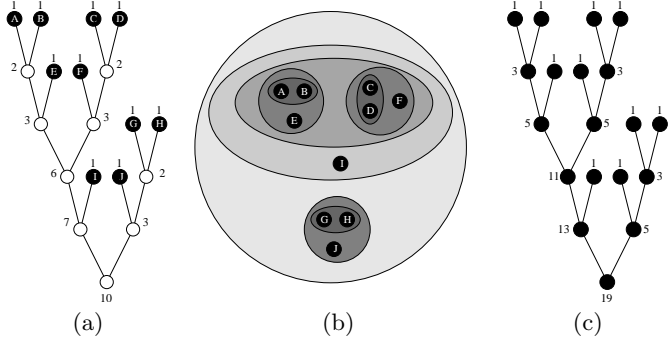}}
\centerline{(a)\hspace{0.33\columnwidth}(b)\hspace{0.33\columnwidth}(c)}
\caption{Community size distribution and analogy with river
networks. (a) Calculation of community sizes from the binary tree. (b)
Representation of the hierarchical structure of nested
communities. (c) Calculation of the drainage area distribution for a
river network.}
\label{area}
\end{minipage}
\end{figure}
The first quantity that will be considered is the community size
distribution. Figure \ref{area}a represents a hypothetical tree
generated by the community identification algorithm (for clarity, the
tree is represented {\it upside down}). Black nodes represent the
actual nodes of the original graph while white nodes are just
graphical representations of groups that arise as a result of the
splitting procedure. Indeed, nodes $A$ and $B$ belong to a community
of size 2, and together with $E$ form a community of size
3. Similarly, $C$, $D$ and $F$ form another community of size 3. These
two groups together form a higher lever community of size 6. Following
up to higher and higher levels, the community structure can be
regarded as the set of nested groups depicted in Figure \ref{area}b.
A natural way of characterising the community structure is to study
the community size distribution. In Figure \ref{area}a, for instance,
there are three communities of size 2, three communities of size 3,
one community of size 6, one community of size 7, and one community of
size 10. Note that a single node belongs to different communities at
different levels.

\begin{figure}[!ht]
  \centerline{\includegraphics[width=0.7\columnwidth]{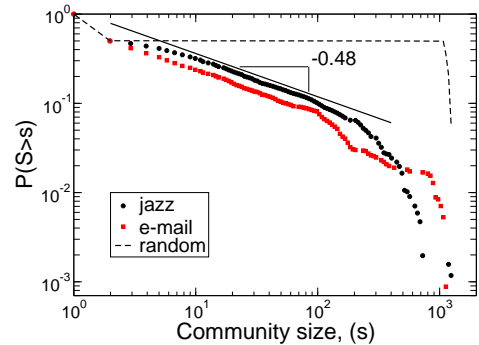}}
  \vspace*{8pt}
  \caption{\label{mail_jazz} Cumulative community size distribution
    $P(S>s)$ as a function of community size $s$ for the email and
    jazz musician networks. The results for the e-mail network are
    plotted in full triangles, while full circles correspond to the
    jazz musicians network. The dotted line corresponds to the results
    obtained in a random network with the same degree distribution as
    the musicians network.}
\end{figure}

Figure \ref{mail_jazz} displays the heavily skewed cumulative
distribution of community sizes, $P(s)$ for both the email network and
the Jazz musicians network.  A comparison of the shape of $P(s)$ shows
a surprising similarity. In both cases, a slow, power law decay with
exponent $0.48$ is observed for community sizes up to $s \sim
200$. This is followed by a faster decay and a cutoff at $s \sim 1000$
corresponding to the size of the systems (the e-mail network
containing 1133 nodes and the jazz network 1265 nodes).  For
small values of $s$ the jazz network deviates from this behaviour,
reflecting the fact that musicians are already grouped in bands of a
certain size, an effect not present in the e-mail network.

The power law of the above distribution suggests that there is no
characteristic community size in the network (up to size 200). To rule
out the possibility that this behaviour is due to our procedure we also
considered the community size distribution for a random graph with the
same size and degree distribution as the e-mail network. In this case
(dotted line in Figure \ref{mail_jazz}), $P(s)$ shows a completely
different behaviour, with no communities of sizes between 10 and 600,
as indicated by the plateau in Figure \ref{mail_jazz}. This corresponds
to a situation in which all the branches (communities) are quite small
(of size less than 10) with the backbone of the network formed by the
union of all these small branches.

\begin{figure}[!ht]
  \centerline{\includegraphics[width=0.7\columnwidth]{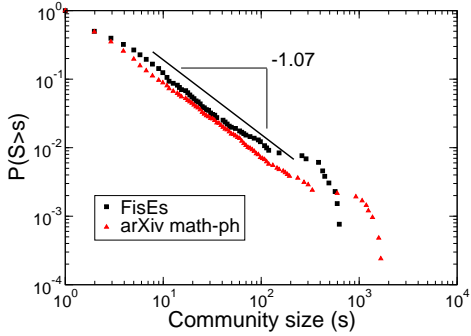}}
  \vspace*{8pt}
  \caption{\label{fises_math} Cumulative community size distribution
    $P(S>s)$ as a function of community size $s$ for the FisEs and
    arXiv math-ph networks. The results for the Fises network are
    plotted in full squares, while full triangles correspond to the
    arXiv mathematical physics network. The full line is shown as a
    guide to the eye and follows a power law with exponent
    $-1.07$. Both distributions fit well to this line, up until $\sim
    1000$ nodes where there is a sharp cutoff corresponding to the
    size of the system (784 nodes for FisEs and 2117 for math-ph).}
\end{figure}

Surprisingly, other networks studied show a power law distribution of
community sizes with a different exponent. In Figure \ref{fises_math}
we see that the exponent is very close to $-1$.

\begin{figure}[!ht]
  \centerline{\includegraphics[width=0.7\columnwidth]{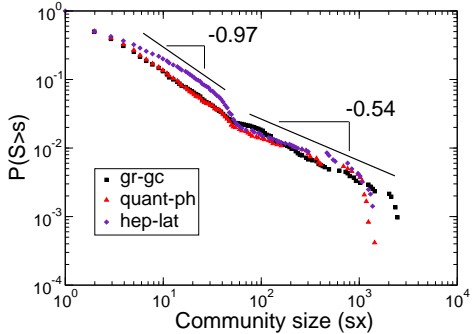}}
  \vspace*{8pt}
  \caption{\label{arxiv_all} Cumulative community size distribution
    $P(S>s)$ as a function of community size $s$ for the arXiv gr-qc,
    quant-ph and hep-lat networks. The results for the gr-qc network
    are plotted in full squares, full triangles correspond to the
    quant-ph network and diamonds represent the hep-lat network. The
    full lines are shown as a guide to the eye and follow power laws
    with exponent $-0.97$ and $-0.54$. Also, all three distributions
    show a sharp cutoff corresponding to the size of the system (2546
    nodes for gr-qc, 1460 for quant-ph and 1411 for hep-lat).}
\end{figure}

More surprising still is the distribution of community sizes in other
arXiv networks. In Figure \ref{arxiv_all} we can see a clear crosover
from one scaling relation to another. All three distributions roughly
follow a power law with exponent $\sim -1$ for community sizes up to
$60$ nodes, whereas between $60$ and $\sim 1000$ nodes the exponent is
seen to be $\sim -0.5$.

\subsection{Analogy with river networks}
Figure \ref{mail_jazz} presents a striking similarity with the
distribution of community sizes and the distribution of drainage areas
in river networks
\cite{rinaldo93,rodriguez96,maritan96,banavar99}. This similarity can
be understood by considering how this distribution is obtained from
the community identification binary tree. Let us assign, as shown in
Figure \ref{area}a, a value of 1 to all the leaves in the binary tree
or, in other words, to all the nodes that represent single nodes in
the original network (black nodes of the binary tree). Then, the size
of a community $i$, $s_i$, is simply the sum of the values $s_{j_1}$
and $s_{j_2}$ of the two communities (or individual nodes), $j_1$ and
$j_2$, that are the offspring of $i$. Figure \ref{area}c shows how the
drainage area of a given point in a river network is
calculated. Consider that at any {\it node} of the river network there
is a source of 1 unit of water (per unit time). Then, the amount of
water that a given node drains is calculated exactly as the community
size for the community binary tree, but adding the unit corresponding
to the water {\it generated} at that point:
$s_i=s_{j_1}+s_{j_2}+1$. This quantity represents the amount of water
that is generated upstream of a certain node. In this scenario, the
community size distribution would be equivalent to the drainage area
distribution of a river where water is generated only at the leaves of
the branched structure.

The similarity between the community size distribution of the e-mail
and jazz networks and the area distribution of a river network is
striking (see, for instance, the data reported in \cite{maritan96} for
the river Fella, in Italy). The exponent of the power law region is
very similar: according to \cite{rinaldo93},
$\alpha_{river}=-0.43\pm0.03$, while for the community size
distribution we obtain $\alpha=-0.48$. Moreover, the behaviour with
first a sharp decay and then a final cutoff is also shared. River
networks are known to evolve to a state where the total energy
expenditure is minimised \cite{kramer92,rinaldo93,sinclair96}. The
possibility that communities within networks might also spontaneously
organise themselves into a form in which some quantity is optimised is
very appealing and deserves further investigation.

\subsection{Horton-Strahler index}
\label{sec:HS}

\begin{figure}
\begin{minipage}[]{\columnwidth}
\centerline{\includegraphics*[width=0.6\columnwidth]{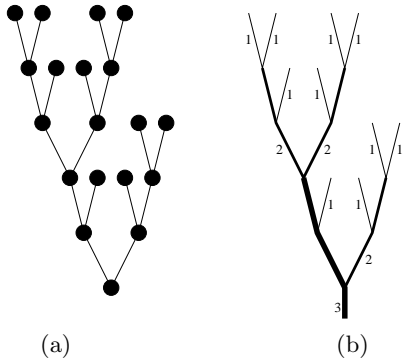}}
\centerline{(a)\hspace{0.4\columnwidth}(b)}
\caption{Calculation of the Horton-Strahler index. (a) Asymmetric
binary tree (b) Corresponding Horton-Strahler indices of the leaves
and branches. In this case there are $N_1=10$ branches with index 1,
$N_2=3$ with index 2 and $N_3=1$ with index 3.}
\label{hsrog}
\end{minipage}
\end{figure}

The similarity between the community size distribution and the
drainage area distribution of river networks prompts one question: is
this similarity arising just by chance or are there other emergent
properties shared by community trees and river networks?  To answer
this question we consider a standard measure for categorising binary
trees: the Horton-Strahler (HS) index, originally introduced for the
study of river networks by Horton \cite{horton45}, and later refined
by Strahler \cite{strahler52}. Consider the binary tree depicted in
the left side of Figure \ref{hsrog}. 
The leaves of the tree are
assigned a Strahler index $i=1$. For any other branch that ramifies into two
branches with Strahler indices $i_1$ and $i_2$, the index is
calculated as follows:
\begin{eqnarray*}
i=\left\{ \begin{array}
{l@{\quad \quad}l}
i_1 + 1       & \mbox{if}\quad i_1 = i_2,\\
\max(i_1,i_2) & \mbox{if}\quad i_1 \neq i_2.
\end{array} \right.
\label{HS}
\end{eqnarray*}
Therefore the index of a branch changes when it meets a branch with
higher index, or when it meets a branch with the same value and both
of them join forming a branch with higher index (see \ref{hsrog}b).

The number of branches $N_i$ with index $i$ can be determined once the
HS index of each branch is known . The bifurcation ratios $B_i$ are
then defined by $B_i=N_i / N_{i+1}$ (by definition $B_i\ge{2}$). When
$B_i\approx B$ for all $i$, the structure is said to be topologically
self-similar, because the overall tree can be viewed as being
comprised of $B$ sub-trees, which in turn are comprised of $B$ smaller
sub-trees with similar structures and so forth for all scales. River
networks are found to be topologically self similar with $3<B<5$
\cite{halsey97}.

\begin{figure}
\vspace*{0.3cm}
\centerline{\includegraphics*[width=0.6\columnwidth]{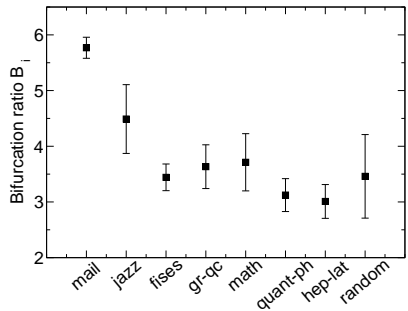}}
\caption{The Horton-Strahler bifurcation ratios $B_i$ and their respective errors.}
\label{hs}
\end{figure}
We find that the community trees seen in Section
\ref{sec:applications} are topologically self similar with
$3<B_i<5.76$ (see Figure \ref{hs}). The same analysis for the
communities in a random graph shows that topological self similarity
does not hold, since the values of $B_i$ are not constant; they
fluctuate more wildly around $3.46$.

The HS index also turns out to be an excellent measure to assess the
levels of complexity in networks. First, let us consider the
interpretation of the index in terms of communities within an
organisation as represented by the email network. The index of a
branch remains constant until another segment of the same magnitude is
found. In other words, the index of a community changes when it joins
a community of the same index. Consider, for instance, the lowest
levels: individuals ($i=1$) join to form a group (with $i=2$), which
in turn will join other groups to form a {\it second level} group
($i=3$). Therefore, the index reflects the {\it level} of aggregation
of communities. For example, in URV one could expect to find the
following levels: individuals ($i=1$), research teams ($i=2$),
departments ($i=3$), faculties and colleges ($i=4$), and the whole
university ($i=5$).  Strikingly, the maximum HS index of the community
tree is indeed 5, as shown in Figure \ref{hs}.

Figure \ref{arbre}b shows the community tree of the e-mail network
with different colours for different HS indices. This helps to
distinguish the individual, team and department levels within a
branch. Actually, the {\it university level} is the ``backbone'' of
the network along which the separation of communities occurs (from the
top to the bottom of the figure). From this backbone, colleges,
departments and some research teams separate, although it is worth
noting that colleges or, in general, centres which are small and have
no internal structure will be classified with a HS index corresponding
to a department or even a team. Therefore, the HS index does not
represent administrative hierarchy but organisational complexity. For
comparison Figure \ref{arbre}c shows in colour the HS index for the
binary tree of a random graph.

The fact that the community structure is topologically self-similar
means that the organisation is similar at different levels. In other
words, it means that individuals form teams in a way that resembles
very much the way in which teams join to form departments, to the way
in which departments organise to form colleges, and to the way in which
the different colleges join to form the whole university.

\section{Conclusions}
The study presented here reveals a characteristic scaling of the
community size distribution of different social networks. The scaling
found follows a power law with two different exponents observed for
different networks. The presence of this particular type of scaling
suggests that some optimising mechanism is responsible for the
self-organisation of social networks. What this mechanism is, remains
to be seen.

\section*{Acknowledgements}
This work was funded by DGES of the Spanish Government (Grant
No. BFM2003-08258) and EC-Fet Open Project IST-2001-33555.
L. D. ackgnowledges the financial support of the Generalitat de
Catalunya (FI2002-00414) and 
P.M.G. that of Fundaci\'on Antorchas.
The authors thank Mark Newman for providing
the arXiv data.

\end{document}